\documentstyle[12pt,psfig,epsf]{article}
\newcommand{\be}{\begin{equation}}
\newcommand{\ee}{\end{equation}}

\newcommand{\beqa}{\begin{eqnarray}}
\newcommand{\eeqa}{\end{eqnarray}}

\newcommand{\eqref}[1]{(\ref{#1})}

\def\boxit#1{\vbox{\hrule\hbox{\vrule\kern8pt
\vbox{\hbox{\kern8pt}\hbox{\vbox{#1}}\hbox{\kern8pt}}
\kern8pt\vrule}\hrule}}
\def\mathboxit#1{\vbox{\hrule\hbox{\vrule\kern8pt\vbox{\kern8pt
\hbox{$\displaystyle #1$}\kern8pt}\kern8pt\vrule}\hrule}}

\def\IB{\relax\hbox{$\inbar\kern-.3em{\rm B}$}}
\def\IC{\relax\hbox{$\inbar\kern-.3em{\rm C}$}}
\def\ID{\relax\hbox{$\inbar\kern-.3em{\rm D}$}}
\def\IE{\relax\hbox{$\inbar\kern-.3em{\rm E}$}}
\def\IF{\relax\hbox{$\inbar\kern-.3em{\rm F}$}}
\def\IG{\relax\hbox{$\inbar\kern-.3em{\rm G}$}}
\def\IGa{\relax\hbox{${\rm I}\kern-.18em\Gamma$}}
\def\IH{\relax{\rm I\kern-.18em H}}
\def\IK{\relax{\rm I\kern-.18em K}}
\def\IL{\relax{\rm I\kern-.18em L}}
\def\IP{\relax{\rm I\kern-.18em P}}
\def\IR{\relax{\rm I\kern-.18em R}}
\def\IZ{\relax\ifmmode\mathchoice
{\hbox{\cmss Z\kern-.4em Z}}{\hbox{\cmss Z\kern-.4em Z}}
{\lower.9pt\hbox{\cmsss Z\kern-.4em Z}} {\lower1.2pt\hbox{\cmsss
Z\kern-.4em Z}}\else{\cmss Z\kern-.4em Z}\fi}

\def\II{\relax{\rm I\kern-.18em I}}

\begin{document}

\hfill  NRCPS-HE-03-36

\vspace{5cm}
%\begin{titlepage}
%\title{
\begin{center}
{\LARGE Tensionless Strings:\\
Physical Fock Space and Higher Spin Fields

}%title ends

\vspace{2cm}
%\author{

{\sl G.K.Savvidy\\
Demokritos National Research Center\\
Ag. Paraskevi, GR-15310 Athens, Hellenic Republic\\
\centerline{\footnotesize\it E-mail: savvidy@inp.demokritos.gr}
}%author ends
%}
%\date{}%in order NOT to write the date
%\maketitle
\end{center}
\vspace{60pt}

\centerline{{\bf Abstract}}

\vspace{12pt}

\noindent
I study the physical Fock space of the tensionless string theory
with perimeter action, exploring its new gauge symmetry algebra.
The cancellation of conformal anomaly requires the space-time to be 13-dimensional.
All particles are massless and there are no tachyon states in the spectrum.
The zero mode conformal operator defines the levels of the
physical Fock space. All levels can be classified
by the highest Casimir operator W of the
little group E(11) for massless particles in 11-dimensions.
The ground state is infinitely degenerated
and contains massless gauge fields of arbitrary large integer spin, realizing
the irreducible representations of E(11) of fixed helicity.
The excitation levels realize CSR representations of little group E(11)
with an infinite number of helicities. After inspection of the first
excitation level, which, as I prove, is a physical null state,
I conjecture that all excitation levels are physical null states.
In this theory the tensor field of the second rank
does not play any distinctive role and therefore one
can suggest that in this model there is no gravity.

%\end{abstract}
%\thispagestyle{empty}
%\end{titlepage}

\newpage

\pagestyle{plain}
%\pagenumbering{roman}

\section{{\it Introduction}}

A string model which is based on the concept of surface
perimeter or length was suggested in  \cite{geo}.
It describes random surfaces embedded in D-dimensional
space-time  with the following action
\be\label{funaction}
S =m \cdot L= {m\over \pi} \int d^{2}\zeta
\sqrt{h}\sqrt{K^{ia}_{a}K^{ib}_{b}},
\ee
where $m$ has dimension of mass, $h_{ab}$ is the induced metric and
$K^{i}_{ab}$ is the second fundamental form
(extrinsic curvature).
There is no Nambu-Goto area term in this action.
Because the action has
dimension of length $L$, alternative to the area,
we have introduced a new dimensional  parameter $m$
which is very similar to the mass parameter
in the action for relativistic point particle.
This observation can be justified better if one takes a limit in which a
surface shrinks into a single world-line, in that case the action
(\ref{funaction}) reduces to its length
\be\label{limit}
S ~~ \rightarrow ~~ m \int ds ,
\ee
and allows to identify a new parameter $m$
with the quantity which is in clear analogy with
the point particle mass.
Geometrically, the action is a natural extension of the
geometrical concept of length, which is well defined for space-time curves,
into two-dimensional surfaces\footnote{In
\cite{polykov} the action is proportional
to the spherical angle and has dimensionless coupling constant.}.

The best way to understand the concept of surface perimeter is to
use simple analogy with general relativity.
In gravity the action measures the
$area$ of the four-dimensional universe alternative to its four-volume.
The action includes scalar curvature term $[R] = cm^{-2}$  and
together with the volume element $[\sqrt{g}d^{4}X ] = cm^4$ has
dimension $cm^2$. The coupling constant in front of the action
should be therefore of  dimension mass square: $m^{2} = 1/G_N$.
In our case dimension of
extrinsic curvature is $[K]=cm^{-1}$ and volume element has dimension
$[d^{2}\zeta \sqrt{h}] = cm^2$, therefore our
coupling constant has dimension of mass.

At the classical level this model is {\it tensionless}
because for the flat Wilson
loop the action is equal to its perimeter $S=m(R+T)$ \cite{geo}.
One can guess therefore that the spectrum of the theory is
massless or even continuous, but because of string rigidity
there are nontrivial vibrational modes and the spectrum
is masseless\footnote{This model
also differs from the tensionless string models based on Schild's work
on "null" string \cite{Schild:vq,deVega:1987hu,Gasperini:1991rv,Lousto:1996hg}
with its most likely continuous spectrum
\cite{Lizzi:1986nv,Lizzi:1989iy,Gamboa:1989zc,Lindstrom:1990qb,
Lindstrom:1990ar,DeVega:1992tm,
Isberg:1992ia,Hassani:rf,Isberg:1993av,
Bonelli:2003kh,Lindstrom:2003mg,Lizzi:1994rn,Zheltukhin:2001jw}.}.
The solution of this nonlinear, high-derivative,
two-dimensional world-sheet CFT is \cite{Savvidy:dv}:
\beqa\label{solution}
X^{\mu}_{L} = x^{\mu} +
{1\over m}\pi^{\mu}\zeta^{+} +
i\sum_{n \neq 0}  {1\over n }~ \beta^{\mu}_{n} e^{-in\zeta^{+}},\nonumber\\
\Pi^{\mu}_{L}=   m e^{\mu} +  k^{\mu}\zeta^{+} + i\sum_{n \neq 0}
{1\over n }~ \alpha^{\mu}_{n} e^{-in\zeta^{+}},
\eeqa
where~$k^{\mu}$ is momentum operator and
$\alpha_n$,$\beta_n$ are oscillators with the
following commutator relations $[x^{\mu},
k^{\nu}] = i\eta^{\mu\nu},~[\alpha^{\mu}_{n},
\beta^{\nu}_{k}]=n~\eta^{\mu\nu}\delta_{n+k,0}$.
The similar expansion exists also for the right
moving modes $\tilde{\alpha}_n$,$\tilde{\beta}_n$ .
The essential difference between tensionless and standard
string theories is the appearance of
additional zero mode $[e^{\mu},\pi^{\nu}]=i\eta^{\mu\nu}$,
meaning that the wave function is  a function of two continuous
variables $k^{\mu}$ and $e^{\mu}$:~ $\Psi_{Phys}= \Psi(k,e)$.
It was suggested in \cite{Savvidy:dv} that $e^{\mu}$
should be interpreted as a polarization vector.
This fact has very important physical consequences. In particular
it leads to {\it high degeneracy of all physical states}.

Our aim is to study spectral properties of the
model and the structure of its physical Fock space
in analogy with the standard string theory
\cite{Brower:1971qr,DelGiudice:1971fp,Brower:ev,
Brower:1972wj,Goddard:iy,Goddard:1972ky}.
This is quite possible using the above mode expansion of the fields and exploring
corresponding new gauge symmetries. In particular one should be
able to prove that the physical Fock space is positive definite,
i.e. ghost-free.

This tensionless theory is invariant under new extended symmetries
\cite{Savvidy:dv}. It is invariant with respect to the
standard conformal transformation
with generators
$$
L_n =\sum_{l}:\alpha_{n-l} \cdot\beta_{l}:
$$
and with respect to the infinite-dimensional
Abelian transformation (\ref{newsymmetry}),(\ref{newsymmetry1})
generated by the operator
$\Theta = \Pi^2 -m^2$ with the following components (\ref{secondcon}),(\ref{constcompo}),(\ref{spectrum}):
$$
k^2,~~~ k \cdot e,~~~ k \cdot \alpha_n,~~~ k\cdot
\tilde{\alpha_n},~~~ \Theta_{nl} .
$$
Therefore in covariant quantization scheme
the space-time equations which define physical Fock space have
the following form \cite{Savvidy:dv}:
\be\label{first}
\left( \begin{array}{l}
  k^2\\
  k\cdot e\\
  k\cdot \alpha_n\\
  k\cdot \tilde{\alpha_n}\\
  \Theta_{nl}\\
  L_n   \\
  \tilde{L_n}
\end{array} \right)\Psi_{Phys}=0,~~~~~~~n,l = 0,1,2,....
\ee
As one can clearly see from the first equation in (\ref{first})
{\it all particles, with arbitrary large spin, are massless}.
This pure massless spectrum is consistent with the tensionless character
of the theory and, what is very important, also tells us that
{\it there are no tachyon states}.

Our aim is to describe physical meaning of different operators in this
system and analyze solutions of these space-time equations in details. They
are similar to the Virasoro equations in Gupta-Bleuler quantization
of standard string theory:
$$
(L_0 - 1)\Psi_{Phys} =0,~~~~~~L_n  \Psi_{Phys} =0 ~~~n = 1,2,.....
$$
where the first equation is a generalization of Klein-Gordon equation
because $L_0 = \alpha^{'}k^2 + \hat{N}$ and
$\hat{N} = {1\over 2}\sum_{n \neq 0}
\alpha_{-n}\alpha_n $ is a mass level operator.
The rest of Virasoro equations allow to prove the absence of negative
norm states on every mass level of physical Fock space and require
the space-time to be 26-dimensional: $D_c =26$.

In our previous consideration of the tensionless strings we
found that the critical dimension is $D_c =13$ \cite{Savvidy:dv}.
This follows from the cancellation  of the conformal anomaly.
The contribution to the
central charge from bosonic coordinates is $2 \times {D\over 12} = {D\over 6}$,
which is twice bigger than in the standard bosonic string theory.
The contribution of Faddeev-Popov ghosts
to the central charge remains the same,
therefore the absence of conformal anomaly requires
the space-time to be 13-dimensional.
\\
\\
It is also important that $L_0$
in our case has different meaning, because it has
the form
$$
L_0 = k \cdot \pi /m ~+ ~\hat{\Xi},
$$
where $\hat{\Xi} =  \sum_{n \neq 0}~ \alpha_{-n} \beta_n$. We shall
define $\hat{\Xi}$ as our new level operator
with eigenvalues $\Xi = 0,1,2...$ and shall identify
the operator $k \cdot \pi /m$ with the length of the
highest Casimir operator of the Poincar\'e group.

A natural question arose, whether
the physical Hilbert space
on every level $\Xi$ is positive-definite,
i.e. ghost-free in analogy with the well known No-ghost theorem
\cite{Brower:1971qr,DelGiudice:1971fp,Brower:ev,
Brower:1972wj,Goddard:iy,Goddard:1972ky}
in the standard string theory.
We shall demonstrate that first
two levels $\Xi =0,1$, ground and excitation
states, are well defined and that there are no negative norm
waves. The vacuum state $\Xi =0$ is infinitely degenerated
and contains massless particles of increasing tensor structure
$A^{\mu_1 ,..., \mu_s}(k)~~~s=1,2,...$. After inspection of the first
level $\Xi =1$, which happens to be a physical null state,
we conjecture that all excitation levels
$\Xi =1,2,3,...$ are {\it physical null states}.

 Let us first describe the ground state, ~{\bf $\Xi =0$}.
The wave function $\Psi_{0}(k,e)~ \equiv~ |k,e,0>$~
is defined as $\alpha_n ~|0,k,e>~ =$ $ ~\beta_n|0,k,e>~=~0,$ $~~~n=1,2,.. $
and the system of equations (\ref{first})
reduces to the following four equations \cite{Savvidy:dv}:
\be\label{second}
k^2 ~\Psi_{0} = 0,~~~~e \cdot k ~\Psi_{0}=0,~~~~
(e^2 - 1)~\Psi_{0}=0,~~~{k \cdot \pi \over m} ~\Psi_{0}=
-\Xi ~\Psi_{0} =0 .
\ee
It follows from the third equation
that ~$\Psi_0$ ~is a function defined on a unit sphere
$e^2 =1$, ~~$\eta^{\mu\nu}=(-+...+)$ and can be
expanded in the corresponding bases:
$$
\Psi_{0} ~= ~|0,k,e> ~= ~A(k) ~+~ A^{\mu_1}(k)~ e_{\mu_1} ~+~
A^{\mu_1 , \mu_2}(k)~ e_{\mu_1} e_{\mu_2} +...|0,k>.
$$
We clearly see that because we
have additional variable~ $e_{\mu}$~{\it
the vacuum state in infinitely degenerated.}
The fields $A^{\mu_1 ,..., \mu_s}(k)$ describe massless
particles of increasing tensor structure. They are symmetric
traceless tensor fields because they are harmonic functions
on a sphere $e^2 =1$, and it follows from the
last equation in (\ref{second}), that
$
k_{\mu_1}~A^{\mu_1,...\mu_s}(k) =0.
$
Thus the fields $A^{\mu_1,...\mu_s}$ are: 1)
symmetric traceless tensors of increasing rank $s=0,1,2,...$,
2) divergent free $k_{\mu_1} ~A^{\mu_1,...\mu_s} =0$
and 3) satisfying massless wave equations
$k^2 ~A^{\mu_1,...\mu_s} =0$.~
All the above conditions are sufficient to
describe integer spin fields
\cite{pauli,singh,fronsdal,Ferrara:1998jm,Haggi-Mani:2000ru,Sundborg:2000wp,
Witten:2000,Bengtsson:1983pg,Bengtsson:1983pd,Vasiliev:1999ba,Sezgin:2001zs,
Mikhailov:2002bp,Francia:2002aa}.
Thus the ground state is infinitely degenerated and contains
gauge particles of arbitrary large fixed helicity. Every tensor gauge field
$A^{\mu_1,...\mu_s},~s=0,1,2,..$ appears only once in the spectrum.
The infinite degeneracy of the vacuum
state reflects enhanced symmetry of the model.
In some sense the leading Regge trajectory
degenerates into a single ground state and one can only
speculate that it may describe unbroken
phase of standard string theory when $ \alpha^{'} \rightarrow \infty$
\cite{gross,Amati:1988tn}.

Two remarks are in order. The first remark is connected
with the infinite degeneracy of the
vacuum state. It has been found that statistical system
of random surfaces with the same perimeter
energy functional (\ref{funaction}) can be
formulated on a cubit lattice, and is
equivalent to a spin system with competing
ferromagnetic and anti-ferromagnetic interaction
\cite{wegner}. It is a remarkable fact that this
spin system also has exponential degeneracy
of the vacuum and excitation states of order $2^{N}$,
where N is the size of the lattice.
This degeneracy should be compared with only double
degeneracy of the vacuum state of the
Ising model with area energy functional.

The second remark is connected with the four vacuum equations
(\ref{second}), they actually coincide with the four equations
introduces by Wigner in 1963
\cite{wigner,brink,abbott,hirata,biedenharn,zoller,floratos} in order to
describe the irreducible representations of the little group of
the Poincar\'e group.
Therefore these equations naturally follow
form the tensionless theory, which is formulated in terms
of Lagrangian mechanics and action principle.
They describe ${\bf 0_s}$ representations of
massless particles of fixed helicity. For these representations
the highest Casimir operator $W = (k \cdot \pi /m)^2  = \Xi^2$ of the little
group $E(11)$, which is Euclidean group of SO(11)
rotations and displacements T(11) for massless particles
in 11-dimensions,
is equal to zero: $\Xi =0$. Alternatively, the excitation levels
realize the second class of infinite dimensional representations
${\bf 0(\Xi)}$ for which $\Xi \neq 0$ and
describe states with an infinite number of helicities. As we shall
see in tensionless theory they are physical null states
and therefore do not produce explicit contradiction with the
obvious experimental fact that there are no particles with infinite
number of helicities.

Indeed, the first excitation state {\bf$\Xi =1$} of the closed string
depends on four polarization tensors  $\xi_{\mu\nu}(e)$,
$\omega_{\mu\nu}(e)$, $\zeta_{\mu\nu}(e)$ and $\chi_{\mu\nu}(e)$.
The restriction on these tensors follows from our space-time
equations (\ref{first}). Solving them we shall see that
$\chi_{\mu\nu}(e) =0$ , $\omega^{\mu\nu} =\omega(e)k_{\mu}k_{\nu}
= -\zeta^{\mu\nu}$ and $\xi_{\mu\nu}(e)$ remain arbitrary, thus
\be
\Psi_1 =  [~\xi_{\mu\nu}(e) \alpha^{\mu}_{-1}
\tilde{\alpha}^{\nu}_{-1} + \omega_{\mu\nu}(e) (\alpha^{\mu}_{-1}
\tilde{\beta}^{\nu}_{-1} - \beta^{\mu}_{-1}
\tilde{\alpha}^{\nu}_{-1}) ~] |0,k,e>.
\ee
The norm of this state
is
$
<\Psi_1 | \Psi_1> =-\omega^{*}_{\mu\nu}\omega^{\mu\nu} =
-\vert \omega \vert^2 ~(k^{2})^{2}
$
and therefore is equal to zero!
It is also normal to the ground state $<\Psi_0 |\Psi_1> =0$. Thus
{\it the first excitation is a physical null state} and realizes
the infinite-dimensional representation ${\bf 0(\Xi)}$ with $\Xi =1$.
This consideration allows to make a conjecture that {\it all
excitations are physical null states} and therefore define
gauge parameters of huge gauge group. In this respect new
extended algebra of our space-time equations (\ref{first})
should play a crucial role defining positive
definite metric in physical Fock space on every level.
\\

In the next sections we shall review solution of
our basic world-sheet equations in conformal gauge
together with the  mode expansion of these
fields (\ref{solution}) and
calculation of the critical dimension in path integral formulation
of the theory and shall formulate our space-time equations (\ref{first})
in Gupta-Bleuler quantization scheme \cite{Savvidy:dv}.
I shall also describe extended symmetries of the model
in terms of new $\Theta$ algebra
(\ref{oldalgebra}), (\ref{newalgebra1}) and
(\ref{newalgebra}). It is
an Abelian extension of conformal algebra and equations
(\ref{fullalgebra}) suggest its oscillator representation.
To this author it is not known, whether this algebra has appeared
also somewhere else.

Our main goal is to solve these space-time equations
for the first few levels and to find out the appropriate interpretation of
the corresponding operators. We shall see that the vacuum is infinitely
degenerated and every massless fixed helicity state
appears only once in the physical spectrum and there are no tachyon
states. In this model the tensor field of the second rank
does not play any distinctive role and therefore one
can suggest that in this model there is no gravity.
The inspection of the first level shows that it is a physical
null state defining gauge transformation of the ground state
wave function.

In the last section I shall also present a short review of
the finite- and infinite-dimensional
representations of the  little group of the Poinca\'e group
defined by Wigner \cite{wigner,brink}. These representations
naturally appear in this tensionless string theory on
different levels of the physical Fock space.

\section{\it Equations and symmetries of the Model B}

The gonihedric action (\ref{funaction}) can be represented
in the form \cite{Savvidy:dv}
\begin{equation}\label{conaction}
S= {m\over\pi}\int d^{2}\zeta \sqrt{h}~\sqrt{ \left(\Delta(h)
X_{\mu}\right)^{2}} ~= ~{1\over\pi}\int d^{2}\zeta ~\Pi^{\mu}
~ \sqrt{h} \Delta(h) X_{\mu}  ,
\end{equation}
here ~$h_{ab}$ ~is the world-sheet
metric,~$\Delta(h)= 1/\sqrt{h}~\partial_{a}\sqrt{h}h^{ab}
\partial_{b} $ ~is  Laplace operator and
I shall consider the model {\it B}, in which
two field variables $X^{\mu}$ and $h_{ab}$ are independent.
This is nonlinear, high-derivative, two-dimensional world-sheet
conformal field theory which can be solved exactly .

To get classical equations and  world
sheet energy-momentum tensor one should
compute variation of the action with respect to the
coordinates $X^{\mu}$
\be\label{motioneq}
(I)~~~~~~~~~~~~~~~~~~~~~~~~~~~~~~\triangle (h)~\Pi^{\mu} =0~~~~~~~~
~~~~~~~~~~~~~~~~~~~~~
\ee
and the metric $h^{ab}$
\be\label{tensor}
(II)~~~~~~T_{ab} = \partial_{\{a }\Pi^{\mu}  ~\partial_{b\}} X^{\mu}
~-~~h_{ab}~h^{cd} ~ \partial_{c}~ \Pi^{\mu}~\partial_{d} X^{\mu}~=~0,
\ee
where we have introduced the operator $\Pi^{\mu}$:
$$
\Pi^{\mu} =m {\triangle (h) X^{\mu} \over
\sqrt{(\triangle (h) X^{\mu})^{2}}}.
$$
It has a property very similar to the constraint equation
for a point-like relativistic particle but with the essential difference
that it is a {\it space-like vector},
\be\label{secondcon}
(III)~~~~~~~~~~~~~~~~~~~\Theta ~\equiv ~\Pi^{\mu}\Pi^{\mu}~ -~ m^{2}~=0.
~~~~~~~~~~~~~~~~~~~~~~~~
\ee
The energy momentum tensor is conserved
$
\nabla^{a}~T_{ab} =0
$
and is traceless $h^{ab}T_{ab} =0$, thus we have {\it high-derivative nonlinear
two-dimensional world-sheet conformal field theory}.
Equations (\ref{motioneq}), (\ref{tensor})  and (\ref{secondcon})
completely define the system. We have equation of motion (\ref{motioneq})
together with the constraint equations (\ref{tensor})
and (\ref{secondcon}). We should stress that in
addition to the conformal generators
(\ref{tensor}) we have an enhanced symmetry generators (\ref{secondcon})
responsible for the symmetry transformation\cite{Savvidy:dv}:
\be\label{newsymmetry}
\delta ~(\sqrt{h}~\Delta(h) X^{\mu}) =
\Omega ~\Pi^{\mu},~~~\delta \Pi^{\mu} =0,~~~\delta h =0.
\ee
Indeed the variation of the action is
\be
\delta S = {1\over\pi} \int d^{2}\zeta ~\Pi^{\mu} ~\delta ( \sqrt{h} \Delta(h)
X^{\mu}) = {m^2\over\pi} \int d^{2}\zeta ~\Omega,
\ee
and is a total derivative because $\Omega =\nabla^{a}~ \sqrt{h} ~ \vartheta_{a}$.
This transformation can be partially integrated out if we add the term
$\chi_a ~\sqrt{h} ~\partial^{a}~\Pi^{\mu} $ to the transformation
(\ref{newsymmetry}).
The action is still invariant, because $\Pi^2 -m^2 =0$ and
$\Pi^{\mu} \partial_a  \Pi^{\mu} =0$.
Then
$
\delta ( \sqrt{h} \Delta(h) X^{\mu}) =
\Pi^{\mu}~\nabla^{a}~\sqrt{h} ~ \vartheta_{a} +
\sqrt{h} ~ \vartheta_a~\partial^{a}~\Pi^{\mu} = \nabla^{a}
(~\Pi^{\mu}~\sqrt{h} ~ \vartheta_a~)
$
and we have our new symmetry transformation in the form
\be\label{alternative}
\delta ( \partial_{a} ~ X^{\mu} ) = \Pi^{\mu} ~ \vartheta_{a},~~~~
\delta \Pi^{\mu} =0,~~~\delta h =0.
\ee
One can also check that $\delta T_{ab} = 0$.
As we shall see, the theory avoids many obstacles of high
derivative field theory, thanks to this extended symmetry.

\section{\it Solution in Conformal Gauge}

We can fix the conformal gauge $h_{ab}=\rho\eta_{ab}$ using
reparametrization invariance of the action and represent
it in the form
(see (\ref{conaction}))
\be\label{gaga}
\acute{S} = {m\over\pi} \int d^{2}\zeta \sqrt{\left(\partial^{2}
X \right)^{2}} ~=~{1\over\pi}\int d^{2}\zeta ~\Pi^{\mu}~\partial^{2} X^{\mu},
~~~~~~~~~~~~\Pi^{\mu} = m \frac{\partial^{2}X^{\mu}}
{\sqrt{\left(\partial^{2}X \right)^2}} .
\ee
In this gauge the equations of motion are more simple
\be\label{confequa}
(I)~~~~~~~~\partial^{2}\Pi^{\mu} =0~~~~~~~~~~~~~~~~~~~~~~~~~~~~~~~~
~~~~~~~~~~~~~~~~~~~~~~~~
\ee
and they should be accompanied by the constraint equations (\ref{tensor})
and(\ref{secondcon})
\beqa\label{cons3}
&(II)&~~~T_{ab} = ~\partial_{\{a } \Pi^{\mu} ~\partial_{b\}} X^{\mu}
-~ ~\eta_{ab}~\partial_{c}~ \Pi^{\mu}~\partial^{c} X^{\mu} =
0,~~~~~~~~~~~~~~~\nonumber\\
\nonumber\\
&(III)&~~~\Theta = \Pi^{\mu}\Pi^{\mu} - m^{2}~=0.~~~~~~~~~~~~~~~
\eeqa
In the light cone coordinates
$\zeta^{\pm}=\zeta^{0}\pm\zeta^{1}$ ~~~$T_{ab}$ takes the form
\beqa\label{cons9}
T_{++}=   2~\partial_{+}\Pi^{\mu}
\partial_{+}X^{\mu},\nonumber\\
T_{--}=   2~\partial_{-}\Pi^{\mu}
\partial_{-}X^{\mu},
\eeqa
with trace equal to zero
$
T_{+-}= 0.
$
The conservation of the energy momentum tensor takes the form
$
\partial_{-}T_{++} = \partial_{+}T_{--}=0
$
and  requires that its components are analytic $T_{++}=T_{++}(\zeta^{+})$
and anti-analytic $~T_{--}=T_{--}(\zeta^{-})$ functions. Thus our
system has infinite number of conserved charges.
This residual symmetry can be easily   seen in gauge
fixed action (\ref{gaga}) written in light cone coordinates
$
\int \sqrt{(\partial_{+}\partial_{-}
X^{\mu})^{2}}~ d\zeta^{+}d\zeta^{-},
$
it is invariant under the transformations
$
\zeta^{+} = f({\tilde{\zeta}^{+}}),~~~\zeta^{-} =
g({\tilde{\zeta}^{-}})
$
where $f$ and $g$ are arbitrary functions. Our new symmetry
transformation is
\be\label{newsymmetry1}
\delta (\partial_{a}X^{\mu}) ~=~ \Pi^{\mu}~  \omega_a
,~~~\Pi^{\mu} \rightarrow \Pi^{\mu} .
\ee
The classical equation  is $ \partial^{2}~\Pi^{\mu}=0 $,
therefore $\Pi^{\mu}$ is a function of the form
\be\label{II}
\Pi^{\mu} =m~{\partial^{2} X^{\mu} \over
\sqrt{(\partial^{2} X)^{2}}} = {1\over 2}(\Pi^{\mu}_{L}(\zeta^{+})
+ \Pi^{\mu}_{R}(\zeta^{-})).\label{seconconsol}
\ee
One can find now that
$
\partial_{+}~\partial_{-} ~X^{\mu} = {1\over 2}[\Pi^{\mu}_{L}(\zeta^{+}) +
\Pi^{\mu}_{R}(\zeta^{-})]~\Omega(\zeta^+ ,\zeta^-),
$
where $\Omega(\zeta^+ ,\zeta^-)$ is arbitrary function
of $\zeta^+$ and $\zeta^-$. Thus $X^{\mu}$ is a sum of
inhomogeneous and homogeneous solutions
\be\label{I}
X^{\mu}=
{1\over 2}[X^{\mu}_{L}(\zeta^+) + X^{\mu}_{R}(\zeta^-)] + {1\over 2}\int^{\zeta^+}_{0}
\int^{\zeta^-}_{0}[\Pi^{\mu}_{L}(\tilde{\zeta}^{+}) +
\Pi^{\mu}_{R}(\tilde{\zeta}^{-})]~\Omega(\tilde{\zeta}^+ ,\tilde{\zeta}^-)
d\tilde{\zeta}^+d\tilde{\zeta}^- ,
\ee
where  $X^{\mu}_{L}(\zeta^+)$, $X^{\mu}_{R}(\zeta^-)$
are arbitrary functions of $\zeta^+$ and $\zeta^-$. We have therefore
two left $X^{\mu}_{L}(\zeta^+)$, $\Pi^{\mu}_{L}(\zeta^+)$ and two
right movers $X^{\mu}_{R}(\zeta^-), \Pi^{\mu}_{R}(\zeta^-)$. These degrees
of freedom are twice bigger than in the standard string theory, this is
simply because world-sheet equations here are of the forth order.
The constraints (\ref{cons9}) take the form
\be
T_{++}={1 \over 2}~\dot{\Pi}^{\mu}_{L}(\zeta^{+})~\dot{X}^{\mu}_{L}(\zeta^+),~~~~~
T_{--}={1 \over 2}~\dot{\Pi}^{\mu}_{R}(\zeta^{-})~\dot{X}^{\mu}_{R}(\zeta^-)],
\ee
verifying the  fact that they are indeed functions of
only one light cone variable and are $\Omega$ $independent$.

\section{\it Critical Dimension  $D_c = 13$}
The action (\ref{gaga}) is invariant under the global symmetries
$\delta X^{\mu} =\Lambda^{\mu\nu}X_{\nu} + a^{\mu}$,
where $\Lambda^{\mu\nu}$ is a constant antisymmetric matrix, while
$a^{\mu}$ is a constant. The translation invariance of the action (\ref{gaga})
$\delta_{a} X^{\mu} = a^{\mu}$ results into the conserved momentum
current
\be
P^{\mu}_{a} = \partial_{a}\Pi^{\mu},~~~~~~~\partial^{a}
P^{\mu}_{a} =0,~~~~P^{\mu} = \int P^{\mu}_0 d \sigma
\ee
($\zeta^1 \equiv \sigma$) and Lorentz transformation
$\delta_{\Lambda} X^{\mu} = \Lambda^{\mu\nu}X_{\nu}$ into angular momentum
current
\be\label{Lorentz}
M^{\mu\nu}_{a} = X^{\mu}\partial_{a}\Pi^{\nu} - X^{\nu}\partial_{a}\Pi^{\mu}
+ \Pi^{\mu}\partial_{a}X^{\nu} - \Pi^{\nu}\partial_{a}X^{\mu},~~~~~~~\partial^{a}
M^{\mu\nu}_{a} =0,~~~~M^{\mu\nu} = \int M^{\mu\nu}_0 d\sigma .
\ee
Last two terms in $M^{\mu\nu}_a$ define internal angular momentum current
and we shall find an appropriate physical interpretation for them considering
representations of the Wigner's little group of the Poincar\'e group
in the next sections. They
actually describe helicities of the massless particles. One can also check
that $\delta M^{\mu\nu} =0$ under transformation
(\ref{alternative}),(\ref{newsymmetry1}).

From its definition the momentum density
$
P^{\mu}_{0}(\zeta^{0},\zeta^{1}) =
\partial_{0}\Pi^{\mu}
$
is conjugate to $X^{\mu}(\zeta^{0},\zeta^{1})$ and therefore
$
[X^{\mu}(\zeta^{0},\zeta^{1}),P^{\nu}_{0}(\zeta^{0},\zeta^{'1})]
=  i\eta^{\mu\nu} \delta(\zeta^{1} - \zeta^{'1}).
$
From (\ref{gaga}) we can deduce the propagator
$
<\Pi^{\mu}(\zeta)X^{\nu}(\tilde{\zeta})>$ $={\eta^{\mu\nu}\over 2}
ln (\vert \zeta - \tilde{\zeta} \vert \mu).
$
Using explicit solution (\ref{I}) one can get
$
<\Pi^{\mu}_{R}(\zeta^-)X^{\nu}_{R}(\tilde{\zeta^-})> =
~\eta^{\mu\nu}~ln [( \zeta^{-} - \tilde{\zeta}^{-})\mu)],
<\Pi^{\mu}_{L}(\zeta^+)X^{\nu}_{L}(\tilde{\zeta^+})> =
~ \eta^{\mu\nu}~ln [( \zeta^{+} - \tilde{\zeta}^{+})\mu)].~
$
Now we are in a position to compute the two-point correlation function
of the energy momentum operator:
\beqa
<T~T_{++}(\zeta^{+}) ~T_{++}(\tilde{\zeta}^{+})  >~
={1\over 4} ~{D \over (\zeta^{+} -
\tilde{\zeta}^{+})^4}.
\eeqa
The ghost contribution to the central charge remains the same as for the
standard bosonic string:
$$
<T~T^{gh}_{++}(\zeta^{+}) ~T^{gh}_{++}(\tilde{\zeta}^{+})  >~
=~-{13\over 4} ~{1 \over (\zeta^{+} -
\tilde{\zeta}^{+})^4},
$$
therefore the absence of conformal anomaly requires the space-time
to be 13-dimensional, close to the 11-dimensional
space-time of M-theory \cite{Savvidy:dv}
\be D_c = 13.
\ee This
result can be qualitatively understood if one takes into account
the fact that the field equations here are of the fourth order and
therefore we have two left movers and two right movers of $X$ and
$\Pi$ fields, thus two times more degrees of freedom than in the
standard bosonic string theory $2 \times {D\over 8}= {D\over 4}$.
We shall confirm this calculation also using mode expansion of
conformal generators $L_n$ in the next section.

\section{\it Mode Expansion and Quantization}
Let us now find mode expansion of different
operators in the general solution (\ref{II}),(\ref{I}).
The appropriate boundary condition for {\it closed strings}
is simply periodicity of the coordinates $X^{\mu}(\zeta^{0},\zeta^{1})=
X^{\mu}(\zeta^{0},\zeta^{1} + 2\pi)$. The arbitrary periodic
functions $X^{\mu}_{L}$ and $X^{\mu}_{R}$  can be written
as normal mode expansions in the form \cite{Savvidy:dv}:
\beqa\label{decom}
X^{\mu}_{L} = x^{\mu} +
{1\over m}\pi^{\mu}\zeta^{+} +
i\sum_{n \neq 0}  {1\over n }~ \beta^{\mu}_{n} e^{-in\zeta^{+}},\nonumber\\
X^{\mu}_{R} =   x^{\mu}
+  {1\over m}\pi^{\mu}\zeta^{-} +
i\sum_{n \neq 0}  {1\over n }~ \tilde{\beta}^{\mu}_{n} e^{-in\zeta^{-}},
\eeqa
where
$
X^{\mu}= {1\over 2}(X^{\mu}_{L}(\zeta^+) + X^{\mu}_{R}(\zeta^-)),
$
and in similar manner
$\Pi^{\mu}  = {1\over 2}(\Pi^{\mu}_{L}(\zeta^{+})
+ \Pi^{\mu}_{R}(\zeta^{-}))
$
\beqa\label{decom1}
\Pi^{\mu}_{L}=   m e^{\mu} +  k^{\mu}\zeta^{+} +
i\sum_{n \neq 0}  {1\over n }~ \alpha^{\mu}_{n} e^{-in\zeta^{+}},\nonumber\\
\Pi^{\mu}_{R}=   m e^{\mu} + k^{\mu}\zeta^{-} +
i\sum_{n \neq 0}  {1\over n }~ \tilde{\alpha}^{\mu}_{n} e^{-in\zeta^{-}}.
\eeqa
Substituting the above mode expansion into the basic commutator
$
[X^{\mu}_{L,R}(\zeta^{\pm}),
P^{\nu}_{L,R}(\zeta^{'\pm})]= 2 \pi i\eta^{\mu\nu} \delta
(\zeta^{\pm} - \zeta^{'\pm})
$
gives
$$
[e^{\mu}, \pi^{\nu}]=[x^{\mu}, k^{\nu}] =  i\eta^{\mu\nu},~~~[\alpha^{\mu}_{n},
\beta^{\nu}_{k}]= n~\eta^{\mu\nu}\delta_{n+k,0},
$$
where $\alpha^{\mu}_{0}=k^{\mu},~\beta^{\mu}_{0}=
\pi^{\mu}/m$~are zero modes momentum. We can also deduce a
less trivial commutator \cite{Savvidy:dv}
\be
[\partial_{\pm}X^{\mu}_{R,L}(\zeta^{\pm}),
\Pi^{\nu}_{R,L}(\zeta^{'\pm})]=  -2 \pi i\eta^{\mu\nu} \delta
(\zeta^{\pm} - \zeta^{'\pm}),
\ee
which tells us that $\Pi$ and $E \equiv - \dot{X}$ form
a second pear of canonically conjugate variables of
our higher derivative theory.
All other commutators are equal to zero. It is important
that commutators and momentum operator $\Pi$ are invariant
under transformation (\ref{alternative}),(\ref{newsymmetry1}).

Using above mode expansion we can find the generators of
the Lorentz group $M^{\mu\nu}$ (\ref{Lorentz})
\be M^{\mu\nu} =
J^{\mu\nu} + E^{\mu\nu} + O^{\mu\nu},
\ee
where
\beqa
J^{\mu\nu} &=& x^{\mu}k^{\nu} - x^{\nu} k^{\mu} =
i(k^{\mu} \partial^{\nu}_k - k^{\nu} \partial^{\mu}_k )\nonumber\\
E^{\mu\nu} &=& e^{\mu}\pi^{\nu} - e^{\nu} \pi^{\mu} = i(
e^{\mu} \partial^{\nu}_{e} - e^{\nu} \partial^{\mu}_{e}) \nonumber\\
O^{\mu\nu} &=& 2i~\sum_{n \neq 0} ~{1\over n}~
(\beta^{\mu}_{-n} \alpha^{\nu }_{n} - \beta^{\nu}_{-n} \alpha^{\mu }_{n}
+ \tilde{\beta}^{\mu}_{-n}\tilde{\alpha}^{\nu }_{n} -
\tilde{\beta}^{\nu}_{-n}\tilde{\alpha}^{\mu }_{n})
\eeqa
The unusual term in this expression is $E^{\mu\nu}$, which acts on the
new variable $e^{\mu}$. Indeed, as we explained in the introduction,
the difference between tensionless and standard string
theories is the appearance of additional zero mode
$ [e^{\mu}, \pi^{\nu}] =  i\eta^{\mu\nu} $, meaning that the wave
function is a function of two continuous variables $k^{\mu}$
and $e^{\mu}$, the momentum and polarization, $\Psi_{Phys}=\Psi(k,e).$
This means that one should define not only initial and final
momenta of the tensionless strings, but also the initial and final
directions of the polarization vector.
This fact has very important consequences.
In particular it leads to {\it high degeneracy of
the physical states}. This can be seen if one expands the wave
function in terms of new vector variable $e^{\mu}$:~~ $\Psi_{phys}
= \sum_s \Psi_{\mu_1 , ... ,\mu_s}e^{\mu_1}...e^{\mu_s}$, these
fields $\Psi_{\mu_1 , ... ,\mu_s}$ belong to the same level of the
Fock space $\Xi$.

These fields are nontrivial helicity
eigenstates of the highest Casimir operator $W = w^{2}_{D-3} $
of the Poincar\'e group, which is the square of the D-3
Pauli-Lubanski form $w_{D-3} = k \wedge M =
k \wedge E$,  and defines ${\bf 0_s}$ or
${\bf 0(\Xi)}$ ~representations of the
little group E(11) depending on whether
$W = \Xi^2$ is zero or not. We shall analyze this expansion in the next
sections.

To guarantee that string states are Lorentz multiplets
one can check the commutators
\be
[L_n , M^{\mu\nu}]= 0,~~~~~~~~~~~~~~~[\Theta_{n,m} , M^{\mu\nu}]=0.
\ee

\section{\it Enhanced Gauge Algebra, $\Theta$-Algebra }
Let us now define the Fourier
expansion of our constraint operators, they are
the conformal operators $L_n$ and our new operators $\Theta_{n,m}$
\be
L_{n}  = <e^{in\zeta^+} :P^{\mu}_{L}~\partial_{+}X^{\mu}_{L}: >,~~~~
\Theta_{n,l}  =  <e^{in\zeta^+ + il\zeta^-} :  \Pi^{\mu}~\Pi^{\mu} -m^2  :>
\label{constcompo}
\ee
thus we have \cite{Savvidy:dv}
\beqa\label{fullalgebra}
L_{n} &=&\sum_{l}:\alpha_{n-l} \cdot\beta_{l}:~~~~~
\tilde{L}_{n} =\sum_{l}: \tilde{\alpha}_{n-l} \cdot \tilde{\beta}_{l}:\nonumber\\
\Theta_{0,0} &=& m^2( e^{2} -1)
+ \sum_{n \neq 0}
{1\over 4 n^2}:(\alpha_{-n}~\alpha_{n} +
\tilde{\alpha}_{-n}~\tilde{\alpha}_{n}):\nonumber\\
\Theta_{n,0} &=& {im\over n}~e \cdot \alpha_{n}
-{1\over 4}\sum_{l \neq 0,n}
{1\over (n-l)l}:\alpha_{n-l}\cdot\alpha_{l}:~~~~~~~~~~~n=\pm1,\pm2,..\nonumber\\
\Theta_{0,n} &=&{im\over n}~e \cdot \tilde{\alpha}_{n}
-{1\over 4}\sum_{l \neq 0,n}
{1\over (n-l)l}:\tilde{\alpha}_{n-l}\cdot\tilde{\alpha}_{l}:
~~~~~~~~~~~n=\pm1,\pm2,..\nonumber\\
\Theta_{n,k} &=& -{1\over 2 n k}:\alpha_{n}\cdot\tilde{\alpha}_{k}:
~~~~~~~~~~~~~~~~~~~~~~~~~~~n,k= \pm 1, \pm 2,....
\eeqa
The conformal algebra has here its classical form but
with twice larger central charge
\be\label{oldalgebra}
[L_n ~,~ L_k] = (n-k) L_{n+k} + {D\over 6}(n^3 -n)\delta_{n+k,0}
\ee
and with the similar expression for right movers $\tilde{L}_n$.
The reason that the central charge is twice bigger than in the
standard bosonic string theory $2 \times {D\over 12} = {D\over 6}$
is simply because we have two left and
two right movers of $X_{\mu}$ field, twice bigger degrees of freedom
\footnote{Such doubling of modes is reminiscent of the bosonic part of the
${\cal N}=2$ superstring \cite{Ademollo:1974fc}. In the last model there was an
essential problem in identifying the $Y^{\mu}$ coordinates which are
introduced in addition to the normal coordinates $X^{\mu}$
\cite{Ademollo:1974fc}.
In our model the coordinate field $X$ has simply
two sets of commuting oscillators and the conjugate fields are
described by a separate field $\Pi_{\mu}$.}.

The full extended gauge symmetry algebra of constraints
(\ref{constcompo}) takes the form
\beqa\label{newalgebra1}
~[~L_n  , \Theta_{0,0}] &=& -2n  \Theta_{n,0}~~~~~~~~~~~~~~~~~~~~~~~~~~~~
[\tilde{L}_n  , \Theta_{0,0}] = -2n \Theta_{0,n} \nonumber\\
~[~L_n , \Theta_{k,0}] &=& -(n+k) \Theta_{n+k,0}~~~~~~~~~~~~~~~~~~~
[\tilde{L}_n , \Theta_{k,0}] = -2n \Theta_{k,n}\nonumber\\
~[~L_n , \Theta_{0,k}] &=& -2n\Theta_{n,k}~~~~~~~~~~~~~~~~~~~~~~~~~~~~
[\tilde{L}_n ,  \Theta_{0,k}]
= -(n+k) \Theta_{0,n+k} \nonumber\\
~[~L_{n} , \Theta_{k,l}] &=& -(n+k) \Theta_{n+k,l}~~~~~~~~~~~~~~~~~~~~
[\tilde{L}_n , \Theta_{k,l}]
= -(n+l) \Theta_{k,n+l} \nonumber\\
~[~L_{n} , ~k^2 ~] &=& 0~~~~~~~~~~~~~~~~~~~~~~~~~~~~~~~~~~~~~~
[~\tilde{L}_n ,~ k^2~]
= 0 \nonumber\\
~[~L_{n} , k\cdot e] &=& -{i\over m} ~k\cdot \alpha_n~~~~~~~~~~~~~~~~~~~~~~~~~~
[~\tilde{L}_n , k\cdot e]
= -{i\over m} ~k\cdot \tilde{\alpha}_n \nonumber\\
~[L_{n} , k\cdot \alpha_l] &=& -l~ ~~k\cdot \alpha_{n+l}~~~~~~~~~~~~~~~~~~~~~~~~
[\tilde{L}_n , k\cdot \tilde{\alpha}_l]
= -l~ ~k\cdot \tilde{\alpha}_{n+l} ,
\eeqa
where we have included commutators with the operators ~$~k^2 , ~~k\cdot e ,~~
k\cdot \alpha_l ,~~ k\cdot \tilde{\alpha}_l$ because they are constituents
of the $\tau$ dependent part of the operator $\Theta$ as one
can see from the formula (\ref{spectrum}). Therefore we added them to the
full algebra of gauge operators.
One should stress that it is an essentially Abelian extension
\be\label{newalgebra}
 [\Theta_{n,k} , \Theta_{l,p}] =
0,~~~~~n,k,l,p=0, \pm 1,\pm 2,...
\ee
One can easily check that Jacobi identities  between all
these operators are satisfied, therefore the relations
(\ref{oldalgebra}), (\ref{newalgebra1}) and (\ref{newalgebra})
define Abelian extension of conformal algebra and hint
that there exist symmetries higher than the
conformal algebra. The equations
(\ref{fullalgebra}) suggest its oscillator representation.
To this author it is not known, whether this algebra has appeared
also somewhere else. In the next section we shall analyze the
physical Fock space, new generators play fundamental
role in defining the particle spectrum of this theory.

\section{\it Physical Fock Space}
To define the physical Hilbert-Fock space we should first
impose our new constraint $\Theta = \Pi^2 -m^2$. The reason
is that as we shall see it defines the spectrum of the theory
\cite{Savvidy:dv}. Indeed the last operator
has a linear and quadratic $\zeta^0\equiv \tau$ dependence which in fact
uniquely defines the spectrum of this string theory:
\be\label{spectrum}
(\Pi^2 -m^2) = k^2 ~\tau^2 + 2\{m e \cdot k ~
+ ~ k\cdot\Pi_{oscil} \} \tau + \Pi^{2}_{oscil} +
2m e \cdot \Pi_{oscil}
+ m^2( e^2 -1).
\ee
The first operator diverges quadratically with $\tau $
and the second one linearly. Therefore in order to have
normalizable states in physical Hilbert-Fock space
one should impose corresponding constraints.
We are enforced to define the physical space as
\be\label{newconst}
k^2 ~ \Psi_{phys} =0,~~~e \cdot k ~\Psi_{phys} =0,
~~~k \cdot \alpha_{n}  ~\Psi_{phys} =0,~~~k \cdot
\tilde{\alpha}_{n} ~\Psi_{phys} =0,~~~~~n=1,2,...
\ee
The first equation states that {\it all physical states with different spins
are massless}. This is consistent with the tensionless character
of the theory. The rest of the constraints which come  from the oscillatory
part of the operator $\Theta = \Pi^2 -m^2$ take the form
\beqa\label{newconstraints}
%\Theta_{0,0} \Psi_{phys}&=&0\nonumber\\
%\Theta_{n,0}\Psi_{phys} &=& 0~~~~~~~~~~~~n= 1,2,....\nonumber\\
%\Theta_{0,n}\Psi_{phys} &=&0 ~~~~~~~~~~~~n= 1,2,....\nonumber\\
\Theta_{n,k}\Psi_{phys}  &=&0 ~~~~~~~~~~~~n,k= 0,1,2,....
%{1\over m^2}\Theta_{0,0} \Psi_{phys}  &=&
%\{( e^{2} -1) + \sum^{\infty}_{n=1}
%{1\over 2n}(a^{+}_{n}~a_{n} +
%\tilde{a}^{+}_{n}~\tilde{a}_{n}) \}\Psi_{phys}= 0
%{1 \over m^2}\Theta_{n,0}\Psi_{phys} &=& \{{i\over \sqrt{n}} e \cdot a_{n}
%-{1\over 4} \sum^{n-1}_{l=1}
%{a _{l}~a_{n-l}\over   \sqrt{l(n-l)}}  -{1\over 2} \sum^{\infty}_{l=1}
%{a^{+}_{l}~a_{n+l}\over \sqrt{l(n+l)}}  \}\Psi_{phys} =0\nonumber\\
%{1\over m^2}\Theta_{0,n}\Psi_{phys} &=& \{{i\over \sqrt{n}}e \cdot \tilde{a}_{n}
%-{1\over 4} \sum^{n-1}_{l=1}
%{ \tilde{a} _{l}~\tilde{a}_{n-l} \over \sqrt{l(n-l)} } -
%{1\over 2} \sum^{\infty}_{l=1}
%{\tilde{a}^{+}_{l}~\tilde{a}_{n+l} \over \sqrt{l(n+l)} }\}\Psi_{phys} =0\nonumber\\
%{1\over m^2}\Theta_{n,k}\Psi_{phys}  &=& \{ - {1\over 2\sqrt{nk}} a_{n}~\tilde{a}_{k}
%\}\Psi_{phys} =0 ~~~~~~~~~~~~n,k= 1,2,....
\eeqa
We have to impose the conformal constraints as well
\beqa\label{physicalh}
(L_0 - a)\Psi_{phys} =0&~~~ (\tilde{L}_0 - a)\Psi_{phys} =&0 \nonumber\\
L_n \Psi_{phys} =0&~~~~~~~~~~~~\tilde{L}_n \Psi_{phys}=&0~~~~~~n =1,2..
%\{\sqrt{n}~ {k \cdot b_n \over m} + \sqrt{n} ~\pi \cdot a_n
%+ \sum^{n-1}_{k=1} \sqrt{k(n-k)}  a_{k} \cdot b_{n-k}  \nonumber\\
%&+& \sum^{\infty}_{k=1} \sqrt{k(n+k)}(  a^{+}_{k} \cdot b_{n+k} +
% b^{+}_{k} \cdot a_{n+k}) \}
%\Psi_{phys} =0
\eeqa
together with "level"\footnote{We shall define this concept below.}
matching condition
\be
( L_0 - \tilde{L}_0)\Psi_{phys}= 0.\label{matching}
\ee
Thus the physical Fock space is defined by the equations
(\ref{newconst}),(\ref{newconstraints}),(\ref{physicalh}) and (\ref{matching}).
We shall make considerable efforts to understand and solve these
{\it space-time equations}.

We are now interested  to study
"level by level"
physical space solving the full system
of space-time equations described above. As we just explained the
essential difference with the standard string theory is the
appearance of additional zero modes
$$
[e^{\mu}, \pi^{\nu}]=[x^{\mu}, k^{\nu}] =  i\eta^{\mu\nu}
$$
meaning that the wave function is a function of
the momentum and polarization vectors $k^{\mu}$ and $e^{\mu}$
\be
\Psi_{Phys}= \Psi(k,e). \label{physicalwave}
\ee
The other difference is that despite the fact that we have the same
conformal algebra as in the standard string theory nevertheless there is an
essential difference consisting in the oscillator representation of
conformal operators.
In particular, as we have seen above, the operators ~$L_0$~and
$~\tilde{L}_0$~  do not
define a mass spectrum any more, instead, as we shall see,
they define the helicities of the massless
representations of the little group of Poincar\'e group.
To make these things more clear
it is helpful to separate zero modes $\alpha_{0} $ and
$\beta_0$ from  other high modes. The ~$L_0$~and  $~\tilde{L}_0$~ operators
will take the form
\beqa\label{conformaleq}
\{~\alpha_{0} \beta_0 + \sum_{n \neq 0}
\alpha_{-n} \beta_n - a ~\} ~\Psi_{Phys}~=~
\{~ {k \cdot \pi \over m }  +
\sum^{\infty}_{n=1} \alpha_{-n} \beta_n +
\beta_{-n}\alpha_{n} ~\}~\Psi_{phys}=0,\nonumber\\
\{~\alpha_{0} \beta_0 + \sum_{n \neq 0}
\tilde{\alpha}_{-n} \tilde{\beta}_n - a ~\} ~\Psi_{Phys}~=~
\{~ {k \cdot \pi \over m }  +
\sum^{\infty}_{n=1} \tilde{\alpha}_{-n} \tilde{\beta}_n +
\tilde{\beta}_{-n}\tilde{\alpha}_{n} ~\}~\Psi_{phys}=0,
\eeqa
where the constant $a$ should be tuned to eliminate vacuum oscillations.
After introducing the "level" operators $\hat{\Xi}_L$ and $\hat{\Xi}_R$ as
\be\label{leveldef}
\hat{\Xi}_L =  \sum^{\infty}_{n=1}(~ \alpha_{-n}
\beta_n + \beta_{-n}\alpha_{n}~),~~~~~
\hat{\Xi}_R =  \sum^{\infty}_{n=1} (~\tilde{\alpha}_{-n} \tilde{\beta}_n +
\tilde{\beta}_{-n}\tilde{\alpha}_{n}~),
\ee
the above space-time equations, which involve $L_0 = k \cdot \pi /m
+ \hat{\Xi}_L $ and $\tilde{L}_0 = k \cdot \pi /m
+ \hat{\Xi}_R$, will take the form
\be\label{leveldef1}
{k \cdot \pi \over m }\Psi_{Phys} = - \hat{\Xi}_L  \Psi_{Phys}
 = - \hat{\Xi}_R  \Psi_{Phys}.
\label{leveleq}
\ee In the vacuum state of all oscillators the eigenvalues of
these operators are equal to zero $\Xi_L = \Xi_R =0$, thus we can
define the vacuum as a zero level state, and the equations reduce
to the following one
\be
{k \cdot \pi \over m }~\Psi_{0}
=0.\label{vacuumeq}
\ee
In general, when oscillators are
in high excitation states,  the eigenvalues of these operators are
integer numbers $\Xi_L = \Xi_R = \Xi = 0,1,2...$ and we can use
them to define {\it levels} of our Fock space. In this
theory, with all its massless states, these numbers allow
to identify subsets $R^{\Xi}$ of states in the full Hilbert-Fock
space. At the same time these numbers actually define
irreducible representations ${\bf 0_s}$ or ${\bf 0(\Xi)}$
of the Euclidean little group E(11). Therefore string
states represent an explicit realization of different irreducible
representations of the (D-2)-dimensional
Euclidean group $E(D-2)$ consisting of $SO(D-2)$
rotations in the transverse hyperplane and $D-2$ displacements T(D-2).
\\

Now let us consider the constraint $\Theta_{0,0}$, it has the form
\beqa\label{last0}
\{~m^2( e^{2} -1) + \sum^{\infty}_{n=1}
{1\over 2n^2}(\alpha_{-n}~\alpha_{n} +
\tilde{\alpha}_{-n}~\tilde{\alpha}_{n}) ~\}~\Psi_{phys} =0
\eeqa
and again for the vacuum state it reduces to the equation
\be
( e^{2} -1)~\Psi_{0} =0.
\ee
One should study in great details this Hilbert-Fock space
in order to learn more about the spectrum of the theory
and to prove the absence of the negative norm states.
Let us start from the ground state.

\section{\it Ground state,~ $\Xi =0$}
First of all let us build the table of equations as follows,
on the left hand side we shall show symbolically
all our space-time equations (\ref{newconst}),(\ref{newconstraints}),
(\ref{physicalh}),(\ref{matching}) and
on the right hand side the equations to which they reduce for the
vacuum state
\beqa
\left( \begin{array}{l}
  k^2\\
  k\cdot e\\
  k\cdot \alpha_n\\
  k\cdot \tilde{\alpha_n}\\
  \Theta_{nm}\\
  L_n\\
  \tilde{L_n}
\end{array} \right)\Psi_{Phys}=0,~~~\Rightarrow~~\left( \begin{array}{l}
  k^2\\
  k\cdot e\\
  0\\
  0\\
  ( e^{2} -1)\\
  k \cdot \pi \\
  k \cdot \pi
\end{array} \right)\Psi_{0}=0.
\eeqa
Thus for the ground state~ $\Psi_{0}~ \equiv~ |k,e,0>$~ which is defined as
$$
\alpha_n ~|0,k,e>~ = ~\beta_n|0,k,e>~=~0, ~~~~~n=1,2,..
$$
from the above consideration we have four equations
\be\label{wigner}
k^2 ~\Psi_{0} = 0,~~~~e \cdot k ~\Psi_{0}=0,~~~~
(e^2 - 1)~\Psi_{0}=0,~~~k \cdot \pi ~\Psi_{0}=0 .
\ee
If one changes by hand the first equation in this system
introducing a  mass term $(k^2  + M^2) \Psi_0 =0 $
then it is easy to see that consistency condition, which comprises
in computing all commutators between four operators
in (\ref{wigner}) will require $M^2 =0$. Indeed~ $[~L_0 , ~e\cdot k~] =
[~k \cdot \pi , e\cdot k] = - i k^2 =i M^2 =0$.

The metric signature is $\eta_{\mu\nu}
=(-1,+1,...,+1)$ and the $e^{\mu}$ is a space-like unit vector.
This fact is not accidental because the initial vector
$\Pi_{\mu}$ was a space-like vector (\ref{secondcon}) and
therefore its zero frequency part $e^{\mu}$ (see mode expansion
(\ref{decom1})) inherited this property. Therefore the first three
equations allow to interpret the vector $e_{\mu}$ as a
polarization vector normal to the momentum vector $k_{\mu}$
\cite{Savvidy:dv}. The space-time wave function $\Psi_{0} =
\Psi(k,e)$ depends therefore on two independent variables:
space-time momentum $k^{\mu}$ and polarization vector $e^{\mu}$
(\ref{physicalwave}). The wave function is a function on a unit
sphere $e^2 =1$ and can be expanded in the corresponding bases
\be\label{vacuumwave}
\Psi_{0} = |0,k,e> = \{A(k) + A^{\mu_1}(k)~
e_{\mu_1} + A^{\mu_1 , \mu_2}(k)~ e_{\mu_1}~ e_{\mu_2}
+...\}|0,k>.
\ee
We clearly see that because in this model we have
additional variable~ $e_{\mu}$~{\it the vacuum state in infinitely
degenerated.} Every field $A^{\mu_1 ,..., \mu_s}(k)$ describes
massless particle of increasing tensor structure. It is a {\it
symmetric traceless tensor field} because it is a harmonic
function on a sphere $e^2 =1$. The constrain (\ref{physicalh}),
(\ref{leveleq}),(\ref{vacuumeq})  reduces to the last equation in
(\ref{wigner})
\be\label{suplemen} k \cdot \pi |k,e,0>~ =
i~k_{\mu} {\partial \over \partial e_{\mu}  }|k,e,0> ~=
i~\{k_{\mu_1}  A^{\mu_1}(k)~+ k_{\mu_1} A^{\mu_1 , \mu_2}(k)~
e_{\mu_2} +...\}|0,k>=0
\ee
and it follows from this equation that
\be\label{divergentfree}
k_{\mu_1}~A^{\mu_1,...\mu_s}(k) =0.
\ee
Thus the fields $A^{\mu_1,...\mu_s}$ are: 1) symmetric tensors of
increasing rank $s=0,1,2,...$,~2) traceless in the sense that
$\eta_{\mu_1 ,\mu_2}A^{\mu_1,...\mu_s}=0$,~3) divergent free
$k_{\mu_1} ~A^{\mu_1,...\mu_s} =0$ and 4) satisfying
massless wave equations $k^2 ~A^{\mu_1,...\mu_s} =0$. All
the above conditions are sufficient to describe integer spin
fields \cite{pauli}. In the vacuum we have therefore massless fields
$A^{\mu_1,...\mu_s}(k)$ of integer spin and momentum
vector $k^{\mu}$, they are fixed helicity states
of the little group $E(11)$. Thus the vacuum is infinitely
degenerated and every fixed helicity state
appears only once in the physical spectrum and there are no tachyon
states. What is also important, our tensor fields expansion is
invariant under Abelian gauge transformation
\be
e_{\mu} \rightarrow e_{\mu} + \chi(k) k_{\mu},
\ee
where $\chi(k)$ is an arbitrary function of momentum $k$.

\section{\it First Excited Level~~ $\Xi =1$}
Let us now consider the first excitation of the most general form
\be
|\Psi_1> = [ ~\xi_{\mu\nu}(e) \alpha^{\mu}_{-1}  \tilde{\alpha}^{\nu}_{-1}
+ \omega_{\mu\nu}(e) \alpha^{\mu}_{-1} \tilde{\beta}^{\nu}_{-1} +
\zeta_{\mu\nu}(e) \beta^{\mu}_{-1}  \tilde{\alpha}^{\nu}_{-1} +
\chi_{\mu\nu}(e) \beta^{\mu}_{-1}  \tilde{\beta}^{\nu}_{-1}~]
|k,e,0>,
\ee
where polarization tensors $\xi_{\mu\nu}(e),\omega_{\mu\nu}(e), \zeta_{\mu\nu}(e)$
and $\chi_{\mu\nu}(e)$ are in general a function of the vector $e_{\mu}$.
Using operator algebra of $(\alpha,\beta)$ oscillators one can
see that
$$
\hat{\Xi}_L  |\Psi_1> ~= \hat{\Xi}_R |\Psi_1> =~|\Psi_1>,
$$
thus justifying the fact that it is a level one state $\Xi_L = \Xi_R =1$
(see equation (\ref{leveleq}). The equation (\ref{leveleq}) now takes the form
\be\label{L0}
{1 \over m } ~k \cdot \pi ~|\Psi_1> ~=
{ i\over m } ~k_{\mu}  {\partial \over \partial e_{\mu} } ~ |\Psi_1> ~
= -    |\Psi_1>.
\ee
Let us now consider the rest of the equations for the first level
wave function:
\beqa
\left( \begin{array}{l}
  k^2\\
  k \cdot e\\
  k\cdot \alpha_1\\
  k\cdot \tilde{\alpha_1}\\
  \Theta_{00}\\
  \Theta_{10}\\
  \Theta_{01}\\
  \Theta_{11}\\
  L_1 \\
  \tilde{L}_1
\end{array} \right)|\Psi_1>=
\left( \begin{array}{l}
  k^2\\
  k\cdot e\\
  k\cdot \alpha_1\\
  k\cdot \tilde{\alpha_1}\\
  m^2(e^2 -1) + {1\over 2} (\alpha_{-1} \alpha_1
  + \tilde{\alpha}_{-1} \tilde{\alpha}_1) \\
  i m ~e \cdot \alpha_1\\
  i m ~e \cdot \tilde{\alpha}_1\\
  -{1\over 2} \alpha_1 \tilde{\alpha}_1\\
  \alpha_0 \beta_1 + \beta_0 \alpha_1\\
  \alpha_0 \tilde{\beta}_1 + \beta_0 \tilde{\alpha}_1
\end{array} \right)|\Psi_1>=0.
\eeqa
The equation $\Theta_{00}|\Psi_1>=0$ takes the form
$$
(e^2 -1) \xi^{\mu\nu} + {1\over 2} \zeta^{\mu\nu} +
{1\over 2} \omega^{\mu\nu}=(e^2 -1) \omega^{\mu\nu} +
{1\over 2} \chi^{\mu\nu}= (e^2 -1) \zeta^{\mu\nu} +
{1\over 2} \chi^{\mu\nu}=(e^2 -1) \chi^{\mu\nu}=
0
$$
with the solution $\chi^{\mu\nu} =0,~ \zeta^{\mu\nu} +
\omega^{\mu\nu} =0,~ e^2 -1 =0$ and we are left with the state
\be\label{physstate}
|\Psi_1> =  [~\xi_{\mu\nu}(e) \alpha^{\mu}_{-1}  \tilde{\alpha}^{\nu}_{-1}
+ \omega_{\mu\nu}(e) (\alpha^{\mu}_{-1} \tilde{\beta}^{\nu}_{-1} -
\beta^{\mu}_{-1}  \tilde{\alpha}^{\nu}_{-1}) ~] |k,e,0>.
\ee
The equations $k\alpha_1 |\Psi_1> =  k\tilde{\alpha_1}|\Psi_1>=0$
tell us that
\be\label{transver}
k^\mu \omega_{\mu\nu} = \omega_{\mu\nu} k^\nu =0.
\ee
The equations $\Theta_{10}|\Psi_1>=
\Theta_{01}|\Psi_1>=\Theta_{11}|\Psi_1> = 0$ take the form
\be\label{transver1}
e^\mu \omega_{\mu\nu} = \omega_{\mu\nu} e^\nu =0
\ee
and finally our constraints
$L_1 |\Psi_1> =  \tilde{L_1}|\Psi_1>=0$ are
\be
\omega_{\mu\nu} \beta^{\nu}_0 |k,e,0> ~=
 {1\over m } \omega_{\mu\nu} \pi^{\nu} |k,e,0> ~=
{ i\over m } \omega_{\mu\nu} {\partial \over \partial e_{\nu} } |k,e,0> ~=
0,\nonumber \\
\ee
and
\be\label{onemore}
k^\mu \xi_{\mu\nu} = \xi_{\mu\nu} k^\nu =0.
\ee
Having in mind the equations (\ref{vacuumwave}), (\ref{suplemen})
for the vacuum wave function the first equation reduces to the form
\be\label{transver2}
\omega_{\mu\mu_1} A^{\mu_1 , \mu_2 ,...\mu_s} =0.
\ee
We can also compute the norm of this state:
\be\label{transver3}
<\Psi_1 |\Psi_1> = -2\omega^{*}_{\mu\nu}\omega^{\mu\nu},
\ee
it is remarkable that it is independent on the tensor $\xi_{\mu\nu}$.
Thus we have to analyze solutions of the equations (\ref{L0}), (\ref{transver}),
(\ref{transver1}),(\ref{onemore}) and (\ref{transver2}).
From (\ref{transver1}) it
follows that nonzero components of $\omega_{\mu \nu}$ are only
$~\omega_{00},~ \omega_{0 ,D-1}, ~\omega_{D-1,0}, ~\omega_{D-1,D-1}~$
simply because this tensor should be orthogonal to $D-2$ space-like
mutually orthogonal polarization vectors
\be
e^{\mu}_{1},~ e^{\mu}_{2},.....,e^{\mu}_{D-2},~~~~e_{i}e_{j}= \delta_{ij}
\ee
and the equations (\ref{transver}) tell us that
$\omega_{00}=\omega_{0 ,D-1}=\omega_{D-1,0} = \omega_{D-1,D-1}$, thus
$$
\omega_{\mu\nu} = \omega(e) k_{\mu}k_{\nu}.
$$
It follows now that the norm  (\ref{transver3}) of our first
level physical state (\ref{physstate}) is zero! The only
remaining restriction on polarizations $\xi_{\mu\nu}$~and ~$\omega_{\mu\nu}$
are equations (\ref{onemore})
and (\ref{L0})
\be\label{polysizations}
k^\mu \xi_{\mu\nu} = \xi_{\mu\nu} k^\nu =0,~~~~~~~k_{\mu}
{\partial \over \partial e_{\mu} } ~
\xi_{\lambda\varrho}(e)  = i m ~ \xi_{\lambda\varrho}(e),~~~~~~
k_{\mu}  {\partial \over \partial e_{\mu} } ~\omega(e)
= i m ~\omega(e).
\ee
Because the first level physical wave function (\ref{physstate})
has zero norm, it defines gauge transformation of the ground state wave
function $\Psi_0$. We have to learn how this gauge transformation acts on
the high spin fields $~A^{\mu_1,...\mu_s}$.

\section{\it Enhanced Gauge Invariance}
Let us summarize our findings. The first two levels of our physical
Fock space are: the ground state wave function $\Psi_0$ which uniquely
describes massless states of increasing helicities $~A^{\mu_1,...\mu_j}$
and the first excited state wave function $\Psi_1$ of the form:
\be
\Psi_1 =  [~\xi_{\mu\nu}\alpha^{\mu}_{-1}  \tilde{\alpha}^{\nu}_{-1}
+ \omega_{\mu\nu} (\alpha^{\mu}_{-1} \tilde{\beta}^{\nu}_{-1} -
\beta^{\mu}_{-1}  \tilde{\alpha}^{\nu}_{-1}) ~] |k,e,0>,
\ee
with nonzero tensors $\xi_{\mu\nu}(e) $ , and
$\omega_{\mu\nu} = k_{\mu}k_{\nu}\omega(e)$, and these tensors
are subject to the
constraint (\ref{polysizations}). The norm of the state $\Psi_1$
is equal to zero and it is orthogonal to the ground state wave function
\be
<\Psi_1 | \Psi_1> =0 ,~~~~~~~~~~ <\Psi_0 | \Psi_1> =0.
\ee
Thus $\Psi_1$ is physical null state, and so we can add it to
any physical state with no physical consequences. We could
therefore impose an equivalence relation
\be
\Psi_0 \sim \Psi_0 + \Psi_1 .
\ee
Let us first consider the case when $\omega(e) =0$. Expanding
the tensor $\xi_{\mu\nu}(e)$ in $e$ series
$
\sum_J \xi^{ \mu_1 ...\mu_s}_{\mu\nu} e_{\mu_1}...e_{\mu_s}
$
from equations (\ref{polysizations}) we can get its properties
\be
k^{\mu} ~\xi^{\mu_1 ...\mu_s}_{\mu\nu} =
k^{\nu} ~\xi^{\mu_1 ...\mu_s}_{\mu\nu}=0,~~~~~~~~~
k_{\mu_{s}}~\xi^{\mu_1 ...\mu_{s-1}~\mu_s}_{\mu\nu} =
i m ~\xi^{\mu_1 ...\mu_{s-1}}_{\mu\nu},~~~~~\Xi =1.
\ee
Then using explicit form of the $\Psi_1$ function
$$
\Psi_1 = (\sum_s \xi^{ \mu_1 ...\mu_s}_{\mu\nu} e_{\mu_1}...e_{\mu_s})~
( \sum_s A_{\mu_1 , ... ,\mu_s}e^{\mu_1}...e^{\mu_s} )~
\alpha^{\mu}_{-1}  \tilde{\alpha}^{\nu}_{-1}|0,k>
$$
the equivalence relation $\delta \Psi_0 = \Psi_1$
will imply the gauge transformation of the
tensor fields $A^{\mu_1 ,..., \mu_s}$ of the form
\be
\begin{array}{l}
  \delta A = A \delta \varphi \nonumber\\
  \delta A^{\mu_1} = A^{\mu_1} \delta \varphi  +
  A \delta \varphi^{\mu_1} \nonumber \\
  \delta A^{\mu_1\mu_2} = A^{\mu_1\mu_2} \delta \varphi +
  A^{\{\mu_1} \delta \varphi^{\mu_2\}}  +
  A \delta \varphi^{\mu_1\mu_2} \nonumber\\
  ............
\end{array}
\ee
where
\be
\begin{array}{l}
  \delta \varphi~~~~~ = ~~\xi_{\mu\nu} ~~~~\alpha^{\mu}_{-1}
  \tilde{\alpha^{\nu}}_{-1}  \nonumber\\
 \delta \varphi^{\mu_1}~~~= ~~\xi^{\mu_1}_{\mu\nu} ~~~\alpha^{\mu}_{-1}
   \tilde{\alpha^{\nu}}_{-1}   \nonumber\\
  \delta \varphi^{\mu_1\mu_2}~ =~
  \xi^{\mu_1\mu_2}_{\mu\nu}~~ \alpha^{\mu}_{-1}
  \tilde{\alpha^{\nu}}_{-1}   \nonumber\\
  ..................................
\end{array}
\ee
These gauge parameters have following properties:
\be
k_{\mu_{s}}~\delta\varphi^{\mu_1 ...\mu_{s-1}~\mu_s}  =
i m ~\delta\varphi^{\mu_1 ...\mu_{s-1}} .
\ee
Analogous transformation is generated by the tensor $\omega_{\mu\nu}(e)
=\omega(e) k_{\mu}k_{\nu}$:~~
$\omega(e) = \sum_s \omega^{ \mu_1 ...\mu_s} e_{\mu_1}...e_{\mu_s},$
with the level one condition, $\Xi =1$
$$k_{\mu_{s}}~\omega^{\mu_1 ...\mu_{s-1}~\mu_s}=
i m ~\omega^{\mu_1 ...\mu_{s-1}},
$$ thus in this case
\beqa
\delta \varphi~&=& \omega~~k_{\mu}k_{\nu}~~(\alpha^{\mu}_{-1}
\tilde{\beta}^{\nu}_{-1} -
\beta^{\mu}_{-1}  \tilde{\alpha}^{\nu}_{-1}) \nonumber\\
\delta \varphi^{\mu_1}~ &=&~\omega^{\mu_1} ~k_{\mu}k_{\nu}~(\alpha^{\mu}_{-1}
\tilde{\beta}^{\nu}_{-1} -
\beta^{\mu}_{-1}  \tilde{\alpha}^{\nu}_{-1}) \nonumber\\
...&.&.................
\eeqa

\section{\it Representations of the Little Group E(11)}

All Lorentz transformations L which leave a fixed null vector
$k_{\mu}=k(1,0,0,1)$ invariant $L~k =k$
form a subgroup called "little group"
\cite{wigner}\footnote{For simplicity we
shall consider in details only four-dimensional space-time.
The extension to high dimension can be found in \cite{brink}, see also
\cite{abbott,hirata,biedenharn,zoller,floratos}.}.
The group of Lorentz transformations which
leave a null vector $k_{\mu}$ invariant is a
two-dimensional Euclidean group of rotations
$R(\theta)=\exp{(-iM_{xy}\theta)}$ in the plane
transverse to the vector $\vec{k}$
and displacements $T^{'}(\alpha)=\exp{(-i\alpha \pi^{'})}$,
$T^{''}(\beta) =\exp{(-i\beta \pi^{''})}$, which are
induced by Lorentz generators
\be
M_{xy},~~~\pi^{'}= M_{zx} - M_{tx},~~~\pi^{''}= M_{zy} - M_{ty}.
\ee
They form the Euclidean  algebra $E(2)$
\be
[\pi^{'},\pi^{''}]=0,~~~[M_{xy},\pi^{'}]=
i\pi^{''},~~~[M_{xy},\pi^{''}]= -i\pi^{'}.\label{little}
\ee
Two Casimir operators of the Poincar\'e group are given by
the operators $P=k^2$ and $W = w^2$ - square of the Pauli-Lubanski vector
$w^{\mu} = \epsilon^{\mu\nu\lambda\rho} k_{\nu}~ M_{\lambda\rho}$
\be
W = {\pi^{'}}^2 + {\pi^{''}}^2
\ee
where we have used
$$w_0 = M_{xy},~~w_z = -M_{xy},~~
w_y = M_{zx} - M_{tx}= \pi^{'},~~ w_x = M_{zy} - M_{ty}=\pi^{''}.$$

To describe representations of the little group one can
take $M_{xy}$ in a diagonal form with integer  eigenvalues
$s=0,\pm 1,\pm 2,...$ and then from
Lorentz subalgebra (\ref{little}) it follows
that both $\pi$ generators have Jacoby:
form \cite{wigner}
\be
M_{xy} = \left( \begin{array}{l}
  .\\
  -2 \\
  ~~-1 \\
  ~~~~~~~~0\\
  ~~~~~~~~~+1\\
  ~~~~~~~~~~~+2\\
  ~~~~~~~~~~~~~~~.
\end{array} \right),~~~
\pi^{'} = \left( \begin{array}{l}
  .\\
  ~~~0~~~~~~~\Xi/2~~~~~~~~~~~~~~~~~~~~~~~\\
  ~~~\Xi/2~~~~~0~~~~~~\Xi/2~~~~~~~~~~~~~~~~~~~~~~~\\
  ~~~~~~~~~~\Xi/2~~~~~0~~~~~~\Xi/2\\
~~~~~~~~~~~~~~~~~~~~\Xi/2,~~~~~0~~~~~~\Xi/2\\
~~~~~~~~~~~~~~~~~~~~~~~~~~~~~\Xi/2,~~~~~0~~~~~~\\
~~~~~~~~~~~~~~~~~~~~~~~~~~~~~~~~~~~~~~~~~~~~~~~.
\end{array} \right),
\ee
\be
\pi^{''} = \left( \begin{array}{l}
  .\\
  ~~~0~~~~~~~i\Xi/2~~~~~~~~~~~~~~~~~~~~~~~\\
  ~~-i\Xi/2~~~~~0~~~~~~i\Xi/2~~~~~~~~~~~~~~~~~~~~~~~\\
  ~~~~~~~~~-i\Xi/2~~~~~0~~~~~~i\Xi/2\\
~~~~~~~~~~~~~~~~~~~~-i\Xi/2,~~~~~0~~~~~~i\Xi/2\\
~~~~~~~~~~~~~~~~~~~~~~~~~~~~~-i\Xi/2,~~~~~0~~~~~~\\
~~~~~~~~~~~~~~~~~~~~~~~~~~~~~~~~~~~~~~~~~~~~~~~~~~~.
\end{array} \right). \label{matrices}
\ee
These infinite-dimensional representations can be
characterized by the parameter $\Xi$
which can be assumed to be any positive real number.
One can now compute the Casimir operator W for
these representations:
\be\label{replittlegroup}
W = {\pi^{'}}^2 + {\pi^{''}}^2 = \Xi^2 .
\ee
Let us define the state with fixed helicity  $s~~(0,\pm 1,\pm 2,...)$ as~~
$M_{xy}|s> =s |s>$. It can be represented as an
infinite vector with entry one
in the row s:
\be
|s> = \left( \begin{array}{l}
  .\\
  0 \\
  1\\
  0\\
  0\\
  0\\
  0\\
  .
\end{array} \right)
\ee
The action of the $\pi^{'}$ generators  on this vector can be found
by using (\ref{matrices})
$$
\pi^{'} |s> = {\Xi\over 2}(|s-1> + |s+1>),~~~
\pi^{''} |s> = {i\Xi\over 2}(|s-1> - |s+1>) .
$$
These relations tell us that under the
Lorentz boosts the state with helicity $|s>$ transforms
into the states with helicities $|s\pm 1> $ with the amplitude
proportional to ${\Xi/ 2}$.

Only in the case when $\Xi =0$ we shall have invariant pure helicity
states $|s>$ which decouple from each other. These are well known
physical representations ${\bf 0_s }$ appearing in
standard field theory of massless particles, where $W =\Xi^2 =0$.

When $\Xi \neq 0$ we shall have representations,
in which helicity of the "particle" can take any integer value
$s~=0,\pm 1,\pm 2,...$. If one prepares a "particle" in
a pure helicity state $s=N$ then after Lorentz boosts
one can find the same "particle" in
different helicity states with nonzero amplitude. These are so called
"continuous spin representations - ${\bf 0(\Xi) }$ " CSR \cite{wigner}.
Let us introduce the coherent state of different helicities
$$
|\varphi> = \sum_{s} e^{i s \varphi} |s> .
$$
Under rotation $R(\theta)$ in xy-plane it will transform as
$$
R(\theta)|\varphi> = \sum_{s} e^{i s \varphi} R(\theta)|s> =
\sum_{s} e^{i s \varphi} e^{in \theta}|s> = |\varphi + \theta>
$$
and under Lorentz boosts as
\be
\pi^{'} |\varphi> = \sum_{s} e^{i s \varphi} \pi^{'}|s> =
\sum_{s} e^{i s \varphi} ~{\Xi\over 2} ~(|s-1> + |s+1>) =
\Xi \cos{\varphi} |\varphi>
\ee
and $\pi^{''} |\varphi> = \Xi \sin{\varphi} |\varphi> $.

In high dimensions the displacement generators are
$\pi_i ,~~i=1,...,D-2$ and the little group for the massless
particles in D-dimensions is a
(D-2)-dimensional Euclidean group
$E(D-2)$ induced by the following Lorentz generators:
\be
M_{ij},~~~\pi_{i} =  M_{D-1 i} - M_{0 i},~~~~i=1,...,D-2
\ee
with commutators
\beqa\label{biglittle}
[~\pi_{i},~~~\pi_{j}~]&=&0,\nonumber\\~[~M_{ij},~\pi_{k}~]&=&
i(\delta_{ik} \pi_{j} - \delta_{jk} \pi_{i}),\nonumber\\
~[M_{ij}, ~M_{kl}]&=& i(\delta_{ik} M_{jl} -\delta_{il} M_{jk}
+ \delta_{jl} M_{ik} - \delta_{jk} M_{il}).
\eeqa
The highest Casimir operator of the Poincar\'e algebra is given by
$W = w^{2}_{D-3}$ - the square of the $D-3$ Pauli-Lubanski form
$w^{\mu_1 ,...\mu_{D-3} }_{D-3} = {1\over \sqrt{2(D-3)!}}
\epsilon^{\mu_1 ,...\mu_{D-3},\nu\lambda\rho}
k_{\nu}~ M_{\lambda\rho}$ and has the form
\be
W = \sum_{i} \pi^{2}_{i}
\ee
If $W =\Xi^2 =0$ we have again finite-dimensional irreducible
representations of the transverse group $SO(D-2)$ which we shall
identify as pure "generalized helicity" states ${\bf 0_s }$.
By generalized helicity one should understand an irreducible representation
with $SO(D-2)$ quantum numbers. If on the
other hand $W =\Xi^2 \neq 0$ the irreducible representations
of the little group $E(D-2)$ are infinite-dimensional and
contain infinite number of irreducible representations of $SO(D-2)$,
and I suggest to denote them again as ${\bf 0(\Xi)}$.

Discussing the representations of the Poincar\'e group Wigner was arguing that
the relativistic wave function should depend on two variables $x$ and $\xi$
in order to describe irreducible representations of the little group.
For that he suggested three equations
(equations (6.5),(6.6)and (6.7) in \cite{wigner}):
\beqa\label{wig1}
(p \cdot p)\psi &=& M^2 \psi \nonumber\\
(\xi \cdot \xi) \psi &=& \psi \nonumber\\
(p \cdot \xi)\psi &=&0,
\eeqa
(in massless case $M^2 =0$) and postulated that the Lorentz generators
should have the form
$$
M^{\mu\nu} = i(p^{\mu} \partial^{\nu}_p
- p^{\nu} \partial^{\mu}_p +
\xi^{\mu} \partial^{\nu}_{\xi} - \xi^{\nu} \partial^{\mu}_{\xi} )
= J^{\mu\nu} + E^{\mu\nu}.
$$
Computing the Casimir operator $W=w^2 = (p \wedge M)^2$
of the Poincar\'e group for this representation
of $M^{\mu\nu}$, that is the square of the
Pauli-Lubanski vector $w^{\mu}$,
he has found that $W = V^2$, where
$V = ip \partial_{\xi}$. The eigenvalue
$\Xi$ of the last operator $V\psi = \Xi \psi$ defines the irreducible
representation of the little group with $W=\Xi^2$.
Thus the  forth Wigner equation is (equation (6.17) in \cite{wigner})
\be\label{wig2}
ip ~{\partial \over \partial \xi}~ \psi = \Xi \psi.
\ee
When $\Xi =0$ the representations describe usual fixed
helicity states of massless particles ${\bf 0_s }$.
When $\Xi \neq 0$
the representations are infinite-dimensional with infinite
helicities  ${\bf 0(\Xi)}$, so called  "continuous spin representations" CSR.

This analysis tells us that our equations (\ref{conformaleq}) and
definition of the "level" operator (\ref{leveldef}),
(\ref{leveldef1}) naturally correspond to the
representation of the little group $E(11)$ with $\Xi_L  = \Xi_R = \Xi$.
In our case $\Xi$ is a level number
and takes only integer values $\Xi =0,1,2,...$.
The ground state corresponds to the representations ${\bf 0_s }$ with
$\Xi =0$ and all high level states of $(\alpha_n,\beta_n)$ oscillators
correspond to the CSR representations ${\bf 0(\Xi)}$ with $\Xi = 1,2,...$.

\section{\it{Conclusion}}
In conclusion let us consider in brief the high levels.
The $\Xi =2$ level contains twenty five tensors because
we have five relevant operators in the left sector
$$
\alpha^{\mu}_{-1}  \alpha^{\nu}_{-1},~~~ \alpha^{\mu}_{-1}\beta^{\nu}_{-1},~~~
\beta^{\mu}_{-1}\beta^{\nu}_{-1},~~~\alpha^{\mu}_{-2},~~~\beta^{\mu}_{-2}
$$
and the same amount in the right sector. Corresponding tensors
are subject to the constraints
\beqa
\left( \begin{array}{l}
  k^2\\
  k \cdot e\\
  k\cdot \alpha_1\\
  k\cdot \tilde{\alpha_1}\\
  k\cdot \alpha_2\\
  k\cdot \tilde{\alpha_2}\\
  \Theta_{00}\\
  \Theta_{10}\\
  \Theta_{20}\\
  \Theta_{01}\\
  \Theta_{02}\\
  \Theta_{12}\\
  \Theta_{21}\\
  \Theta_{11}\\
  \Theta_{22}\\
    L_1 \\
  \tilde{L}_1\\
   L_2 \\
  \tilde{L}_2
\end{array} \right)|\Psi_2>=
\left( \begin{array}{l}
  k^2\\
  k\cdot e\\
  k\cdot \alpha_1\\
  k\cdot \tilde{\alpha_1}\\
  k\cdot \alpha_2\\
  k\cdot \tilde{\alpha_2}\\
  m^2(e^2 -1) + {1\over 2} (\alpha_{-1} \alpha_1
  + \tilde{\alpha}_{-1} \tilde{\alpha}_1) + {1\over 8} (\alpha_{-2} \alpha_2
  + \tilde{\alpha}_{-2} \tilde{\alpha}_2)\\
  i m ~e \cdot \alpha_1 + {1\over 8} \alpha_{-1} \alpha_2\\
  {i m\over 2} ~e \cdot \alpha_2 - {1\over 4} \alpha_{1} \alpha_1\\
  i m ~e \cdot \tilde{\alpha}_1 + {1\over 8} \tilde{\alpha}_{-1} \tilde{\alpha}_2)\\
  {i m\over 2} ~e \cdot \tilde{\alpha}_2 -
  {1\over 4} \tilde{\alpha}_{1} \tilde{\alpha}_1\\
  -{1\over 4} \alpha_1 \tilde{\alpha}_2\\
  -{1\over 4} \alpha_2 \tilde{\alpha}_1\\
    -{1\over 2} \alpha_1 \tilde{\alpha}_1\\
    -{1\over 8} \alpha_2 \tilde{\alpha}_2\\
  \alpha_0 \beta_1 + \beta_0 \alpha_1 +\alpha_{-1} \beta_2 +
  \beta_{-1} \alpha_2\\
  \alpha_0 \tilde{\beta}_1 + \beta_0 \tilde{\alpha}_1 + \tilde{\alpha}_{-1}
  \tilde{\beta}_2 +   \tilde{\beta}_{-1} \tilde{\alpha}_2\\
  \alpha_2 \beta_0 + \beta_2 \alpha_0 + \alpha_1 \beta_1\\
  \alpha_0 \tilde{\beta}_1 + \beta_0 \tilde{\alpha}_1 + \tilde{\alpha}_{1}
  \tilde{\beta}_1
\end{array} \right)|\Psi_2>=0.
\eeqa
It is rather complicated system of space-time equations and we
shell describe solution of these equations in a separate
place.

I should also mention that in our early attempts
\cite{Savvidy:1997vg,Savvidy:1998sb} to
write down space-time equations for the tensionless strings we were
using the general finite and infinite-dimensional representations of the
Lorentz group defined as $(j,\lambda)$, where j is integer or
half-integer spin and $\lambda$ is arbitrary complex number, very
similar to the parameter $\Xi$ introduced in the previous section. If
$\lambda$ is real and is equal to $j$ plus natural number $N=1,2,3..$
then representation $(j,\lambda)$ is finite-dimensional,
otherwise it is infinite-dimensional and contains all representations
of the little group with spins $j+1,j+2,....$. The number of
representations used in our approach was growing polynomially
with spin in order to produce physically acceptable spectrum.
This grows should be compared with the exponential grows of
spin states in Ramond generalization of the Dirac equation \cite{ramond}.

The author wishes to thank D.Gross for hospitality in
KITP in Santa Barbara, where part of this work was completed
and would like also to thank P.Ramond, C.Thorn
and R.Woodard for pointing out the Wigner's articles \cite{wigner}
and hospitality in University of Florida.
This work was supported in part by the
EEC Grant no. HPRN-CT-1999-00161.

\vfill
\end{document}